\documentclass[twocolumn]{aastex631}
\usepackage[normalem]{ulem}

\newcommand{\PSi}{$\rm P_{\text{Si\,\textsc{ii}}}$}

\begin{document}

\title{Line Polarization of Si\,{\sc ii} $\lambda$6355\AA\, in Type Ia Supernovae:\\ A New Statistical Approach to Probe the Explosion Physics and Diversity}

\author[0000-0001-7101-9831]{Aleksandar Cikota}
\affiliation{Gemini Observatory / NSF's NOIRLab, Casilla 603, La Serena, Chile; aleksandar.cikota@noirlab.edu}

\author[0000-0002-4338-6586]{Peter Hoeflich}
\affiliation{Department of Physics, Florida State University, Tallahassee, FL 32306-4350, USA} 

\author[0000-0003-1637-9679]{Dietrich Baade}
\affiliation{European Organisation for Astronomical Research in the Southern Hemisphere (ESO), Karl-Schwarzschild-Str.\ 2, 85748 Garching b.\ M\"{u}nchen, Germany}

\author[0000-0002-0537-3573]{Ferdinando Patat}
\affiliation{European Organisation for Astronomical Research in the Southern Hemisphere (ESO), Karl-Schwarzschild-Str.\ 2, 85748 Garching b.\ M\"{u}nchen, Germany}

\author[0000-0001-7092-9374]{Lifan Wang}
\affiliation{George P. and Cynthia Woods Mitchell Institute for Fundamental Physics and Astronomy, Texas A\&M University, 4242 TAMU, College Station, TX 77843, USA}

\author[0000-0003-1349-6538]{J. Craig Wheeler}
\affiliation{Department of Astronomy, University of Texas, Austin, TX 78712-1205, USA}

\author[0000-0002-6535-8500]{Yi Yang}
\affiliation{Physics Department and Tsinghua Center for Astrophysics, Tsinghua University, Beijing 100084, People's Republic of China}

\author[0009-0001-9148-8421]{Elham Fereidouni}
\affiliation{Department of Physics, Florida State University, Tallahassee, FL 32306-4350, USA}

\author[0000-0001-5965-0997]{Divya Mishra}
\affiliation{Department of Physics and Astronomy, Texas A\&M University, 4242 TAMU, College Station, TX 77843, USA}
\affiliation{George P. and Cynthia Woods Mitchell Institute for Fundamental Physics \& Astronomy, Texas A\&M University, 4242 TAMU, College Station, TX 77843, USA}

\begin{abstract}
Spectropolarimetry provides a unique probe of ejecta asphericities, offering direct insights into the underlying explosion physics of Type Ia supernovae (SNe\,Ia).
We analyze the statistical properties of pre-maximum spectropolarimetric data for 24 SNe\,Ia observed with VLT/FORS, focusing on the Si\,{\sc\,ii} $\lambda$6355\,\AA\, line. 
Previous studies have revealed a correlation between the peak Si\,{\sc\,ii} polarization degree and the expansion velocity. Here, we combine these observations with multi-dimensional non-thermodynamical equilibrium radiative transfer simulations. We consider two asphericity classes: (i) lopsided abundance distributions produced by off-center delayed-detonation transitions in near-$\rm~M_{Ch}$ white dwarfs or, for example, WD collisions (Class~I), and (ii) global, axisymmetric density asphericities such as those arising from explosions of rapidly rotating WDs or mergers (Class~II). Our model grid spans normal to subluminous SNe\,Ia and successfully reproduces the observed Si\,{\sc ii} velocity–polarization trend, with higher velocities associated with stronger asphericities. 
Consistent with observations, transitional SNe\,Ia and the faint end of the normal SNe\,Ia population show the highest Si\,{\sc\,ii} polarization and are best explained by Class~I scenarios. In contrast, subluminous SNe\,Ia are dominated by Class~II asphericities, characterized by lower Si\,{\sc\,ii} polarization but significant continuum polarization. The observed distribution of Si\,{\sc\,ii} polarization depends on both the observer's viewing angle $\theta$ and the intrinsic asphericity. Statistical analysis of these spectropolarimetric snapshots enables the separation of Class~I and Class~II contributions and highlights the intrinsic diversity among SNe\,Ia.
Our results imply viewing-angle-dependent luminosities in our local sample, which may have implications when using high-redshift SNe~Ia as evidence for the need of non-standard cosmology.
\end{abstract}

\keywords{Supernovae --- Polarimetry}

\section{Introduction}
\label{sect:introduction}

Type Ia Supernovae (SNe\,Ia) are thermonuclear explosions of carbon/oxygen white dwarfs. They are standardizable candles \citep{1993ApJ...413L.105P} and, therefore, serve as distance indicators on cosmological scales \citep{1998AJ....116.1009R,1999ApJ...517..565P}.
Although supernovae have been systematically discovered since the 1930s \citep{1934PNAS...20..254B}, the progenitor systems and explosion mechanisms of SNe\,Ia are still highly debated. The main progenitor scenarios are (i) the single-degenerate scenario in which a white dwarf accretes material from a companion star until it reaches the Chandrasekhar mass limit and explodes \citep[e.g.][]{1973ApJ...186.1007W}; (ii) the double-degenerate scenario \citep[e.g.][]{1984ApJS...54..335I, 1984ApJ...277..355W}, in which a white dwarf either accretes mass from another white dwarf or merges with its companion white dwarf to explode; (iii) the core-degenerate scenario \citep[e.g.][]{2003ApJ...594L..93L,2011MNRAS.417.1466K}, in which the white dwarf merges with a core of a giant star during a common envelope phase. Alternative scenarios also suggest(iv) the isolated white dwarf scenario, in which the explosion may be triggered by a thermonuclear runaway \citep{2015MNRAS.448.2100C}; or (v) the triple system scenario, in which two white dwarfs collide \citep[][]{2017MNRAS.465L..44K}.

The explosion mechanism is also debated. In the (i) double-detonation scenario \citep[][]{1982ApJ...257..780N,1982ApJ...253..798N,1994ApJ...423..371W} the explosion starts with a helium-shell detonation at the surface of a sub-Chandrasekhar-mass white dwarf, producing a shock wave that triggers a secondary detonation in the C/O core. In the (ii) delayed-detonation scenario \citep[e.g.][]{1991A&A...245..114K, Hoeflich_1995, 2013MNRAS.429.1156S} the explosion starts with a subsonic deflagration in the core, which later transitions to a thermonuclear detonation once the deflagration wave reaches the upper Chapman–Jouguet point; and in the (iii) violent merger scenario \citep[e.g.][]{Pakmor_2012} two white dwarfs collide, triggering the supernova explosion. See \citet{2018PhR...736....1L}, \citet{2021ApJ...922..186H}, and \citet{2025A&ARv..33....1R} for more detailed reviews of progenitor systems and explosion mechanisms.

Various attempts have been made to distinguish between these possible progenitor systems and explosion mechanisms. These include studying the environments of the SNe (e.g., dust extinction, light echoes, variable Na I D lines, etc.), searching for signatures of companion stars at very early epochs, and investigating ejecta asphericities and explosion geometries. 

Spectropolarimetry is a powerful technique that allows the study of 3D shapes of objects that cannot otherwise be resolved \citep[e.g.][]{1993PASAu..10..263S,Hoeflich_1995,1996ApJ...462L..27W}. The low intrinsic continuum polarization measurements (P$_{\rm cont}$ $\lesssim$ 0.3$\%$) of normal SNe\,Ia imply that the explosions are relatively spherical \citep{Cikota2019MNRAS.490..578C}, while subluminous SNe display continuum polarizations of up to $\sim$ 0.8$\%$, corresponding to asphericities of up to $\sim$20$\%$ when viewed equator-on (e.g.\ SN\,1999by and SN\,2005ke, \citealt{2001ApJ...556..302H} and \citealt{2012A&A...545A...7P}, respectively).

SNe\,Ia display a prominent polarization signal in the absorption lines, particularly the Si\,{\sc ii} $\lambda$6355 \AA\, and the near-IR Ca II triplet lines \citep{2008ARA&A..46..433W}. To first order, line polarization can be understood as arising when the absorber is not spherically distributed in front of the underlying polarized photosphere formed by Thomson scattering \citep{2003ApJ...593..788K,2008ARA&A..46..433W}. Thus, it is a direct consequence of the ejected material located in front of the SN-photosphere. However, this is a first-order approximation only and, in fact, polarization by electrons and spectral lines, including Si\,{\sc ii}, form in the same layers, requiring full radiation-transport calculations (see Fig.\ 15 in \citealt{2023MNRAS.520..560H}).

\citet{1991A&A...246..481H, Hoeflich_1995, 1995ApJ...444..831H, 1996ApJ...459..307H, 2015MNRAS.450..967B,2016MNRAS.455.1060B, 2016MNRAS.462.1039B, 2021ApJ...922..186H, 2023MNRAS.520..560H} predicted different polarization signatures for different explosion mechanisms of SNe\,Ia. Therefore, studying line polarization is among the most promising methods for distinguishing between different explosion mechanisms. 

\citet{Cikota2019MNRAS.490..578C} systematically analyzed a sample of 35 SNe\,Ia, observed at 127 epochs in total with the Focal Reducer and Low-Dispersion Spectrograph (FORS, \citealt{Appenzeller_etal_1998}) on the Very Large Telescope (VLT) between 2001 and 2015. The spectropolarimetric data is used to perform a statistical study of correlations between ejecta geometry and observable parameters.
In \citet{Cikota2019MNRAS.490..578C} we found a Si\,{\sc ii} velocity vs.\ Si\,{\sc ii} peak polarization relationship (see Fig.~\ref{fig:PSiII_vel_asymmetry}), which, we speculated, could be explained by a combination of an off-center delayed-detonation transitions (DDT) model and a brightness decline (see also Fig.\ 13 and Sect.\ 5.3.\ in \citealt{Cikota2019MNRAS.490..578C}). In this work, we further test that hypothesis and support it with simulations of off-center delayed-detonation models, which successfully reproduce the Si\,{\sc ii} velocity vs.\ Si\,{\sc ii} peak polarization relationship.\\

In Sect.~\ref{sect:observations} we present the observed sample of SN\,Ia polarization measurements, in Sect.~\ref{sect:simulations} we describe the model and simulations, and in Sect.~\ref{sect:results} we present and discuss the results, followed by the conclusions and discussion of future directions in Sect.~\ref{sect:summary}.

\section{Sample of Type I\lowercase{a} Supernova observations}
\label{sect:observations}

The sample of observed SNe\,Ia is a subset of the observations presented in \citet{Cikota2019MNRAS.490..578C}. We include all 24 SNe for which we have at least one spectropolarimetric observation taken between -11.0 and +1.0 days relative to B-band peak brightness.

\begin{figure}
\centering
\includegraphics[width=0.5\textwidth]{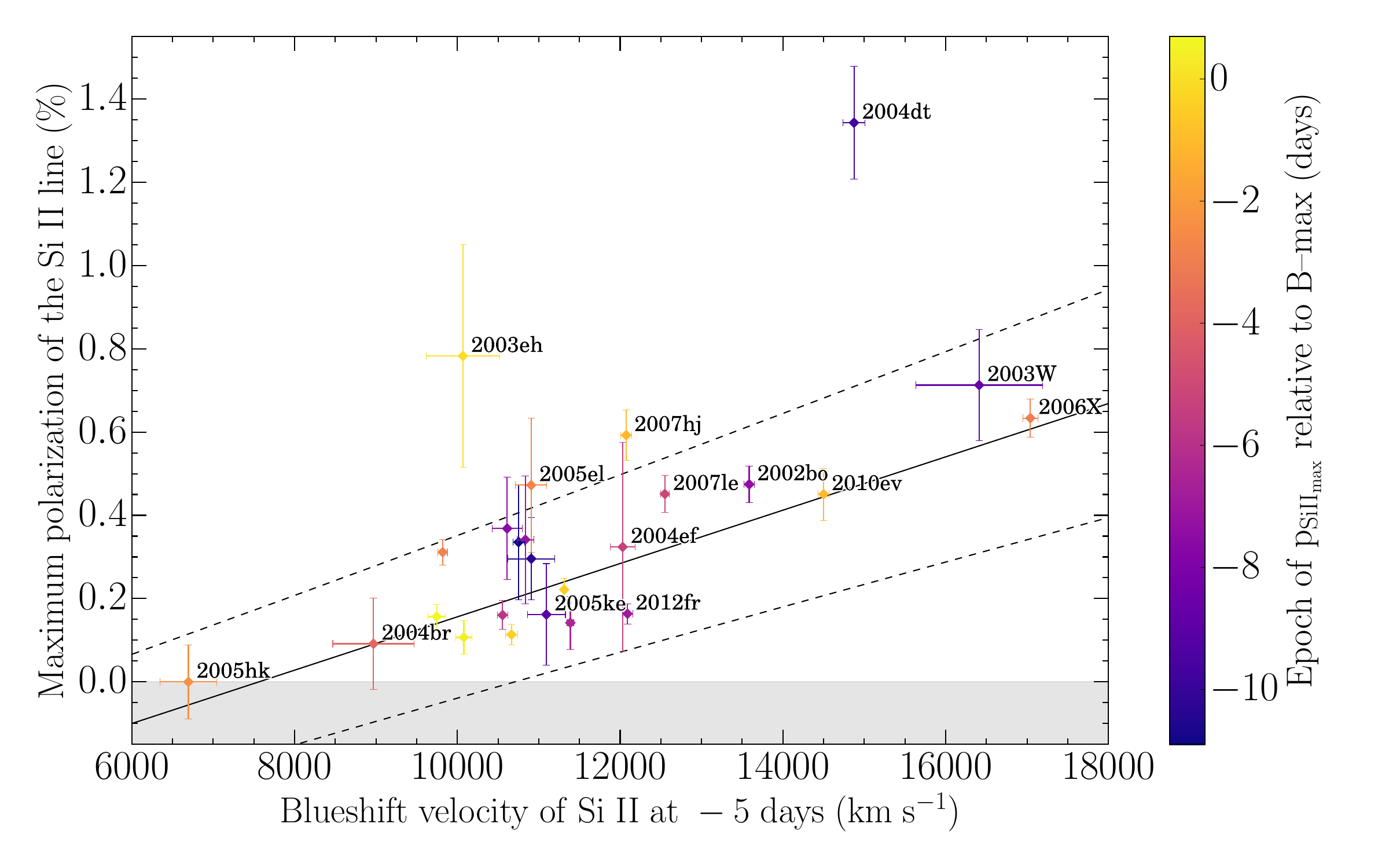}
\vspace{-7mm}
\caption{Maximum linear polarization of the Si\,{\sc ii} $\lambda$6355\,\AA\, line between -11 and +1 days relative to peak brightness, as a function of the interpolated Si\,{\sc ii} $\lambda$6355 \AA\, velocity at -5 days relative to B-band peak brightness. The black solid line shows the linear least-squares fit to the data, while the dashed lines indicate the 1$\sigma$ uncertainty. The figure is reproduced from \citet{Cikota2019MNRAS.490..578C}.}
\label{fig:PSiII_vel_asymmetry}
\end{figure}

Each SN was observed at between 1 and 13 epochs. In \citet{Cikota2019MNRAS.490..578C}, we measured the peak polarization across the Si\,{\sc ii} $\lambda$6355 \AA\, line profile from polarization spectra binned to 100 \AA, plotting the maximum value across all epochs (i.e.\ the highest measured polarization value between -11.0 and +1.0 days relative to B-band peak brightness) in Fig.~\ref{fig:PSiII_vel_asymmetry}, as a function of the Si\,{\sc ii} $\lambda$6355 \AA\, velocity 5 days before B-band peak brightness.
The polarization spectra were corrected for polarization bias\footnote{Polarization bias occurs because the polarization degree $P = \sqrt{Q^2 + U^2}$ is inherently positive, even when the true polarization is zero. As a result, noise in the Stokes parameters $Q$ and $U$ tends to produce an artificial increase in the measured value of $P$.} following \citealt{1997ApJ...476L..27W} (see equation 10 in \citealt{Cikota2019MNRAS.490..578C}).

To determine the Si\,{\sc ii} velocity 5 days before peak brightness, we measured (in \citealt{Cikota2019MNRAS.490..578C}) the photospheric rest-frame wavelengths of the absorption minima in the FORS flux spectra, combined with spectra from the Open Supernova Catalog \citep{2017ApJ...835...64G}, and interpolated the velocity at -5 days relative to peak brightness. In \citet{Cikota2019MNRAS.490..578C}, we found a strong linear relationship (with a Pearson correlation coefficient $\rho \gtrsim$ 0.8) between the degree of linear polarization of the Si\,{\sc ii} $\lambda$6355 \AA\, line and the Si\,{\sc ii} $\lambda$6355 \AA\, line velocity (Fig.~\ref{fig:PSiII_vel_asymmetry}). 

Furthermore, we also determined the $\Delta$m$_{15}$ values by fitting photometric data from the Open Supernova Catalog with SNooPy \citep{2011AJ....141...19B, 2015ascl.soft05023B}. There is a clear trend of lower \PSi\ toward brighter supernovae and larger \PSi\ for underluminous SNe \,Ia. This trend was previously found by \citet{2007Sci...315..212W}, and was later confirmed and discussed in \citet{2009A&A...508..229P} and \citet{Cikota2019MNRAS.490..578C}.

The measurements are given in Table~\ref{tab:observations}. For details on data reduction, see \citet{Cikota2019MNRAS.490..578C}.

\begin{table}
	\centering
	\tiny
	\caption{\label{tab:observations} The sample of observed Type Ia supernovae is taken from \citealt{Cikota2019MNRAS.490..578C}. The Si\,{\sc ii} velocity is given at -5 days relative to B-band peak brightness. The Si\,{\sc ii} polarization corresponds to the maximum value measured across all epochs.}
	\begin{tabular}{lcccccccc} 
	\hline
SN Name & $\Delta$m$_{15}$ & Si\,{\sc ii} velocity & Si\,{\sc ii} polarization & Pol. epoch \\
 & (B mag) & (km s$^{-1}$) & (per cent) & (days) \\
\hline
SN 2001el & 1.13 $\pm$ 0.04 & 9746 $\pm$ 109 & 0.16 $\pm$ 0.03 & 0.7 \\
SN 2001V & 0.73 $\pm$ 0.03 & 11388 $\pm$ 39 & 0.14 $\pm$ 0.06 & -6.4 \\
SN 2002bo & 1.12 $\pm$ 0.02 & 13585 $\pm$ 66 & 0.47 $\pm$ 0.04 & -7.4 \\
SN 2002el & 1.38 $\pm$ 0.05 & 10611 $\pm$ 186 & 0.37 $\pm$ 0.12 & -7.6 \\
SN 2002fk & 1.02 $\pm$ 0.04 & 10080 $\pm$ 96 & 0.11 $\pm$ 0.04 & 0.4 \\
SN 2003eh & $\dots$ & 10068 $\pm$ 451 & 0.78 $\pm$ 0.27 & 0.0 \\
SN 2003W & 1.30 $\pm$ 0.05 & 16412 $\pm$ 781 & 0.71 $\pm$ 0.13 & -8.7 \\
SN 2004br & 0.68 $\pm$ 0.15 & 8968 $\pm$ 500 & 0.09 $\pm$ 0.11 & -3.9 \\
SN 2004dt & 1.21 $\pm$ 0.05 & 14873 $\pm$ 137 & 1.34 $\pm$ 0.14 & -9.7 \\
SN 2004ef & 1.45 $\pm$ 0.01 & 12031 $\pm$ 151 & 0.32 $\pm$ 0.25 & -5.3 \\
SN 2004eo & 1.32 $\pm$ 0.01 & 10907 $\pm$ 285 & 0.3$\pm$ 0.1 & -10.3 \\
SN 2005cf & 1.18 $\pm$ 0.01 & 10555 $\pm$ 63 & 0.16 $\pm$ 0.03 & -5.8 \\
SN 2005de & 1.41 $\pm$ 0.06 & 10752 $\pm$ 70 & 0.34 $\pm$ 0.14 & -10.9 \\
SN 2005df & 1.06 $\pm$ 0.02 & 9819 $\pm$ 57 & 0.31 $\pm$ 0.03 & -2.8 \\
SN 2005el & 1.40 $\pm$ 0.01 & 10906 $\pm$ 192 & 0.47 $\pm$ 0.16 & -2.7 \\
SN 2005hk & 1.47 $\pm$ 0.14 & 6694 $\pm$ 350 & 0.0 $\pm$ 0.09 & -2.3 \\
SN 2005ke & 1.66 $\pm$ 0.14 & 11093 $\pm$ 234 & 0.16 $\pm$ 0.12 & -9.1 \\
SN 2006X & 1.09 $\pm$ 0.03 & 17041 $\pm$ 90 & 0.63 $\pm$ 0.05 & -3.0 \\
SN 2007hj & 1.95 $\pm$ 0.06 & 12074 $\pm$ 66 & 0.59 $\pm$ 0.06 & -1.1 \\
SN 2007le & 1.03 $\pm$ 0.02 & 12549 $\pm$ 51 & 0.45 $\pm$ 0.04 & -5.1 \\
SN 2008fp & 1.05 $\pm$ 0.01 & 11313 $\pm$ 23 & 0.22 $\pm$ 0.03 & -0.5 \\
SN 2010ev & 1.12 $\pm$ 0.02 & 14500 $\pm$ 69 & 0.45 $\pm$ 0.06 & -1.1 \\
SN 2010ko & 1.56 $\pm$ 0.05 & 10835 $\pm$ 102 & 0.34 $\pm$ 0.15 & -7.1 \\
SN 2011iv & 1.77 $\pm$ 0.01 & 10665 $\pm$ 76 & 0.11 $\pm$ 0.02 & -0.4 \\
SN 2012fr & 0.80 $\pm$ 0.01 & 12090 $\pm$ 64 & 0.16 $\pm$ 0.02 & -6.9 \\
\hline
	\end{tabular}\\
\end{table}

\section{Models and simulations}
\label{sect:simulations}

The final result of a Type Ia supernova explosion depends on initial conditions, including progenitor history, accretion rates, and initial metallicity \citep[e.g.][]{HoeflichKhokolov1996,Hoeflich_2017}. To probe the details of the underlying physics and to constrain the viewing angle of the observer requires a comprehensive analysis, including light curves, the time evolution of flux and polarization spectra with a multi-wavelength approach from the UV, via the optical to the mid-IR \citep{2006NewAR..50..470H,2023MNRAS.520..560H}. 

Instead, we focus on the statistical properties of a sizable SNe Ia sample. Specifically, we examine a particular observable at a given time, using explosion scenarios that have been successfully applied in previous detailed analyzes.

The simulations include scenarios for off-center delayed-detonation (DD) models for core-normal SNe\,Ia, where the deflagration-to-detonation transition (DDT) occurs at a mass ($\rm M_{DDT}$) close to the Chandrasekhar limit ($\rm M_{Ch}$), and sub-$\rm M_{Ch}$ explosions.

\begin{figure*}
\includegraphics[width=1.0\textwidth]{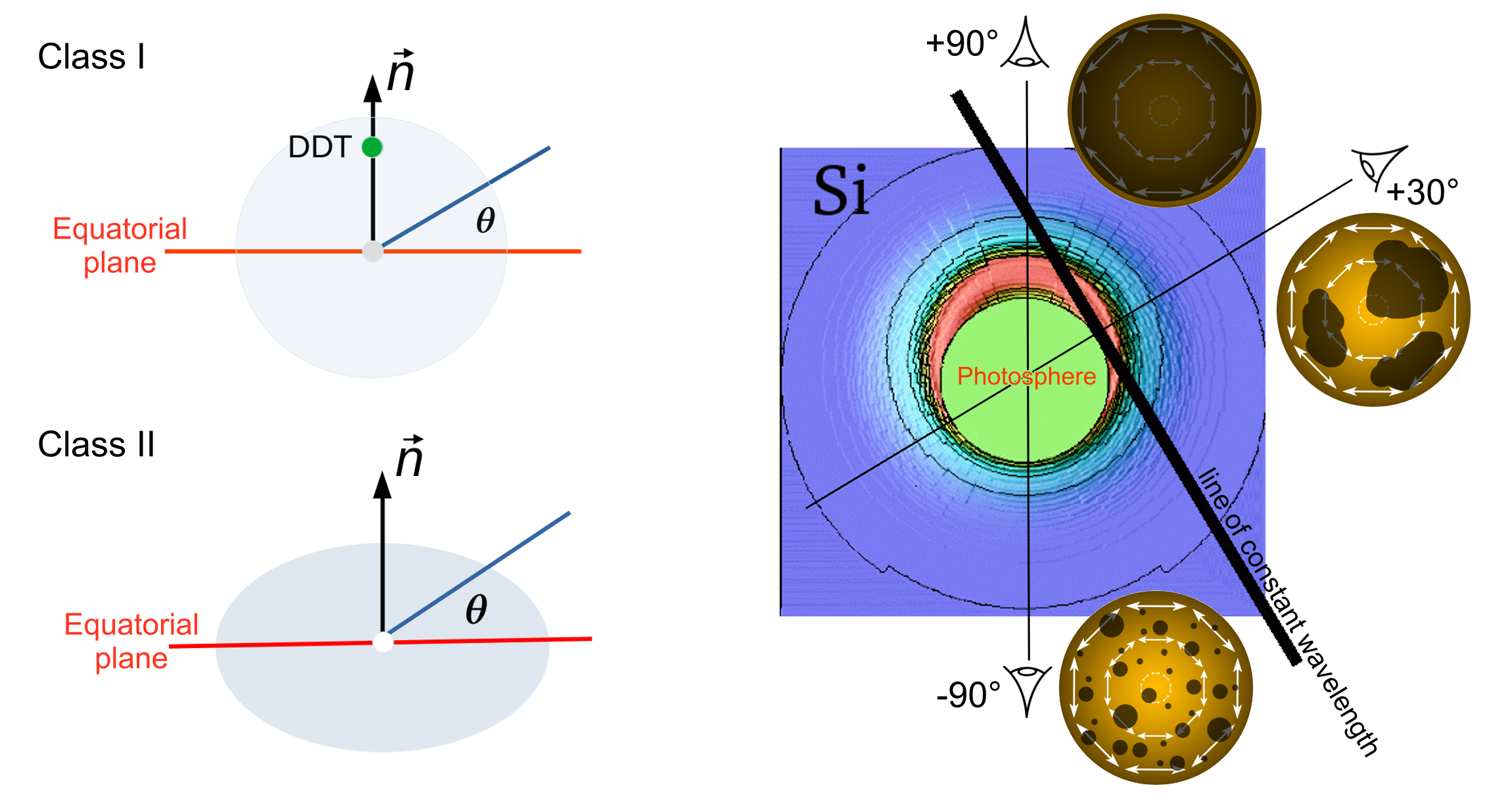}
\caption{\textit{Left:} Illustration of two model classes of asphericities: (i) lopsided large-scale configurations from off-center detonations (Class~I) and (ii) global density asphericities (Class~II), adapted from \citealt{2001ApJ...556..302H}. \textit{Right:} Schematic illustration of polarization formation in an off-center DDT, viewed from different angles (adapted from \citealt{2006NewAR..50..470H}). The measured net polarization at a viewing angle of +90 degrees is zero, owing to the nearly spherical distribution of absorbing material (e.g., Silicon). At +30 degrees, the aspherical distribution produces strong line polarization and pronounced loops in the Stokes $Q$-$U$ plane, whereas at -90 degrees, small clumps produce weaker polarization and smaller loops. 
Any combination with the mechanisms illustrated in the left figures will produce loops in time series in the $Q$-$U$ plane (except in the case of a purely axisymmetric geometry), similar to what has been observed (e.g. \citealt{1997ApJ...476L..27W, Cikota2019MNRAS.490..578C, 2020ApJ...902...46Y}).}
\label{fig:Schematics}
\end{figure*}

There are two model classes: Class~I, representing lopsided large-scale configurations from off-center detonations, and Class~II representing global density asphericities, as illustrated in Fig.~\ref{fig:Schematics}. 

For normal SNe\,Ia, with normal brightness, characterized by large-scale lopsided abundance distributions, we consider models 16, 23, and 25, as described in \citealt{hk96} and \citealt{Hoeflich_2017}. These models represent Class I asphericities. Note that a similar morphology to Class I may be produced by a stellar collision of two WDs and, possibly, sub-$\rm M_{Ch}$ WD, at least in the outer layers considered here. 

Furthermore, we also adopt parameterized rotationally symmetric density and abundance configurations, which represent Class II asphericities. They may be produced by thermonuclear explosions originating from rotating WDs of both $\rm M_{Ch}$ and sub-$\rm M_{Ch}$ or two merging WDs (models 10, HeD10 and HeD6; see \citealt{hk96} and \citealt{Hoeflich_2017}).

The parameter space representing all explosion scenarios and progenitor channels is vast, and simulations are expensive, and observations at a specific time of an SN\,Ia are insufficient to characterize the individual objects. 
The simulations used here provide a representative sample of delayed-detonation scenarios capable of reproducing individual SNe \,Ia with different brightnesses.

The questions to be addressed by these simulations are: (i) Can the classes of asphericity explain the diversity of polarization $\rm P$; (ii) What are the characteristic dependencies on the viewing-angle $\theta$; (iii) is there a spread in the polarization-velocity relation of Si\,{\sc ii} and (iv) is there a dependence of the velocity and polarization of Si\,{\sc ii} on the peak brightness decline relation $\Delta$m$_{15}$. In addition, for off-center DDTs, we briefly discuss the implications for the underlying physics.

\subsection{Numerical Tools and Setup of Simulations} 
\label{sec:numerical_tools}

The simulations were performed with our multidimensional non-LTE HYDrodynamical RAdiation \textit{HYDRA} code that consists of modules for the solution of nuclear reaction networks, the particle and photon transport scheme, and detailed nuclear and atomic networks (\citealt{1995ApJ...443...89H,2021ApJ...923..210H,2023MNRAS.520..560H,2024JPhCS2742a2024H,2025arXiv250107654H}, and references therein). Polarization spectra can be calculated using our Monte Carlo scheme through post-processing of given explosion models and atomic-level populations, since the statistical equations are independent of the orientation of the photon field's electrical vector (e.g.\ \citealt{Hoeflich_1995,2006NewAR..50..470H,2012A&A...545A...7P,2020ApJ...902...46Y,2023MNRAS.520..560H}). In all simulations, we used photon counters with an angular resolution of approximately $\Delta \theta \approx 15^\circ$.

Although complex networks with hundreds of levels per ion are used, several ionization states are particularly important for the optical [Si\,{\sc ii}] multiplet at $\lambda \lambda \lambda$6241, 6248, 6273\,\AA\, with the second quantum state attributed to the 2s, 2d, 2f orbitals with low excitation energy of the low level. During the photospheric phase, the UV transitions to the ground state are optically thick, leading to an overpopulation by a factor of 10 to 100 relative to LTE in the line-forming region and hence to a large blocking of the radiation field from inner layers with a Thomson optical depth of 1. 
Although a very intuitive picture for the formation of polarization (and absorption lines) due to an aspherical distribution of the absorbing material in front of the photosphere was suggested by \citet{1997ApJ...476L..27W}, in reality the polarization-forming region and the line-forming regions coincide (see Figs. 12 \& 15 in \citealt{2023MNRAS.520..560H}, and references therein).

\section{Results}
\label{sect:results}

The simulation results are summarized in Table~\ref{table:simulationsresults}. The models are based on \citealt{HoeflichKhokolov1996} and \citealt{Hoeflich_2017}, and span a range of $\Delta$m$_{15}$ from 1.1 to 1.9 mag in the B band. Model 10 assumes an elliptical oblate density structure with an axis ratio of A/B = 0.82 and reproduces subluminous SNe\,Ia.
Model 16 represents transitional SNe\,Ia, and models 23 and 25 represent normal-bright SNe\,Ia.
Furthermore, the sub-$\rm M_{Ch}$ SNe\,Ia are represented by models HeD10 and HeD6, which include detonating He layers.

\begin{table*}
	\centering
	\tiny
    \setlength{\tabcolsep}{2pt} 
	\caption{\label{table:simulationsresults} Model properties for lopsided chemical distributions produced by an off-center DDT at $\rm M_{DDT}$ (given as a fraction of the WD mass), and asymmetric, oblate elliptical density structures with an axis ratio A/B for DD (model 10) originating from double degenerated WD mergers or a rapidly rotating WD, normal-bright (models 23 and 5), transitional SNe (model 16), and for the sub-$\rm M_{Ch}$ SNe (models HeD10 and HeD6). The models are based on \citealt{HoeflichKhokolov1996} and \citealt{Hoeflich_2017}. Given are the axis ratio A/B or $\rm M_{DDT}$, the expansion velocity v(Si\,{\sc ii}), V($\theta$):=v(Si\,{\sc ii},$\theta$)/v(Si\,{\sc ii}), and $\rm P_{max}$(Si\,{\sc ii}) as a function of the inclination angle $\theta$. 
    }
	\begin{tabular}{lcccccccccccccccccc} 
	\hline 
Model&Type&Remark&$\rm \Delta m_{15,s}(B)$&$\rm \Delta m_{15,s}(V)$& {v(Si\,{\sc ii})}&V(90)&V(60)&V(30)&V(0)&V(-30)&V(-60)&P(90)&P(60)&P(30)&P(0)&P(-30)&P(-60)&\\
     &    &      &(mag)                   &(mag)           &($\rm \times 1000~km~s^{-1}$)&&    &     &    &      &      &     &     &     &    &      &      &       \\

\hline
10&DD&spherical&1.9&1.41&8.5&1&1&1&1& 1& 1&0&0&0&0&0&0                               \\
10&DD&A/B=0.82&1.9&1.41&8.5&1.05&1.05&1.12&1.2& 1.12& 1.05&0&0.1&0.25&0.35&0.25&0.1& \\
&                                                                       \\
&                                                                         \\
16&DD&spherical&1.41&1.18&10.4&1&1&1&1&1 &1 &0&0&0&0&0&0& \\
16&DD&$\rm M_{DDT}$=0.3&1.33&1.08&10.4&1.6&1.45&1.2&1&1 & 1&0&1.2&2.2&0.2&0.05&0\\
16&DD&$\rm M_{DDT}$=0.6&1.43&1.18&10.4&1.3&1.29&1.15&1& 1&1 &0&0.8&1.2&0.7&0.2&0\\
16&DD&$\rm M_{DDT}$=0.9&1.45&1.18&9.8&1.1&1.05&1&1& 1&1 &0&0.1&0.2&0.1&0.05&0\\
&                                                                                   \\
23&DD&spherical&1.21&0.75&13&1&1&1&1& 1&1 &0&0&0&0&0&0\\
23&DD&$\rm M_{DDT}$=0.3&1.21&0.75&13&1.4&1.25&1.05&1&1 & 1&0&0.6&0.5&0.25&0.1&0.05\\
23&DD&$\rm M_{DDT}$=0.6&1.21&0.75&13&1.6&1.45&1&1& 1& 1&0&1.7&1.5&0.7&0.3&0.1\\
23&DD&$\rm M_{DDT}$=0.9&1.21&0.75&13&1.1&1.06&1&1&1 &1 &0&0.6&0.4&0.2&0.13&0.03\\
&                                                            \\
25&DD&spherical&1.1&0.68&13&1&1&1&1&1 & 1&0&0&0&0&0&0\\
25&DD&$\rm M_{DDT}$=0.3&1.1&0.68&13&1.2&1.14&1.04&1& 1&1 &0&0.2&0.2&0.12&0.1&0.05\\
25&DD&$\rm M_{DDT}$=0.6&1.1&0.68&13&1.7&1.43&1.25&1& 1&1 &0&1.1&0.8&0.4&0.2&0.1\\
25&DD&$\rm M_{DDT}$=0.9&1.1&0.68&13&1.2&1.1&1.05&1& 1& 1&0&0.55&0.45&0.25&0.15&0.05\\
&\\
\\
HeD10&&A/B=0.82&1.36&1.03&15&0.95&1.02&1.15&1.2& 1.15& 1.03&0&0.2&0.4&0.5&0.4&0.2\\
HeD10&&A/B=0.68&1.36&1.03&15&0.85&1.02&1.19&1.3& 1.19& 1.03&0&0.4&0.85&1.1&0.85&0.4\\
HeD6&&A/B=0.82&1.48&1.14&11&0.9&1.05&1.15&1.18& 1.15& 1.05&0&0.1&0.2&0.3&0.2&0.1\\
\hline
	\end{tabular}\\
\end{table*}

\subsection{Maximum linear polarization \PSi\ vs.\ Si\,{\sc ii} velocity and $\Delta$m$_{15}$}
\label{sect:max_lin_pol_v_dm15}

In this section, we discuss the maximum linear polarization P in Si\,{\sc ii} $\lambda$6355 \AA\ (measured over time) as a function of the Si\,{\sc ii} $\lambda$6355 \AA\ velocity and the decline rate $\Delta$m$_{15}$ from the simulations and compare them to the observations.

For the models, we assume a decline rate $\rm \Delta m_{15,s}$ measured from the spherical equivalent. To avoid degeneracy at the low brightness end, we applied the stretch $s$ to the time axis. The typical uncertainties are about $\pm$ 0.1 mag.

\begin{figure*}
\centering
\includegraphics[width=0.49\textwidth]{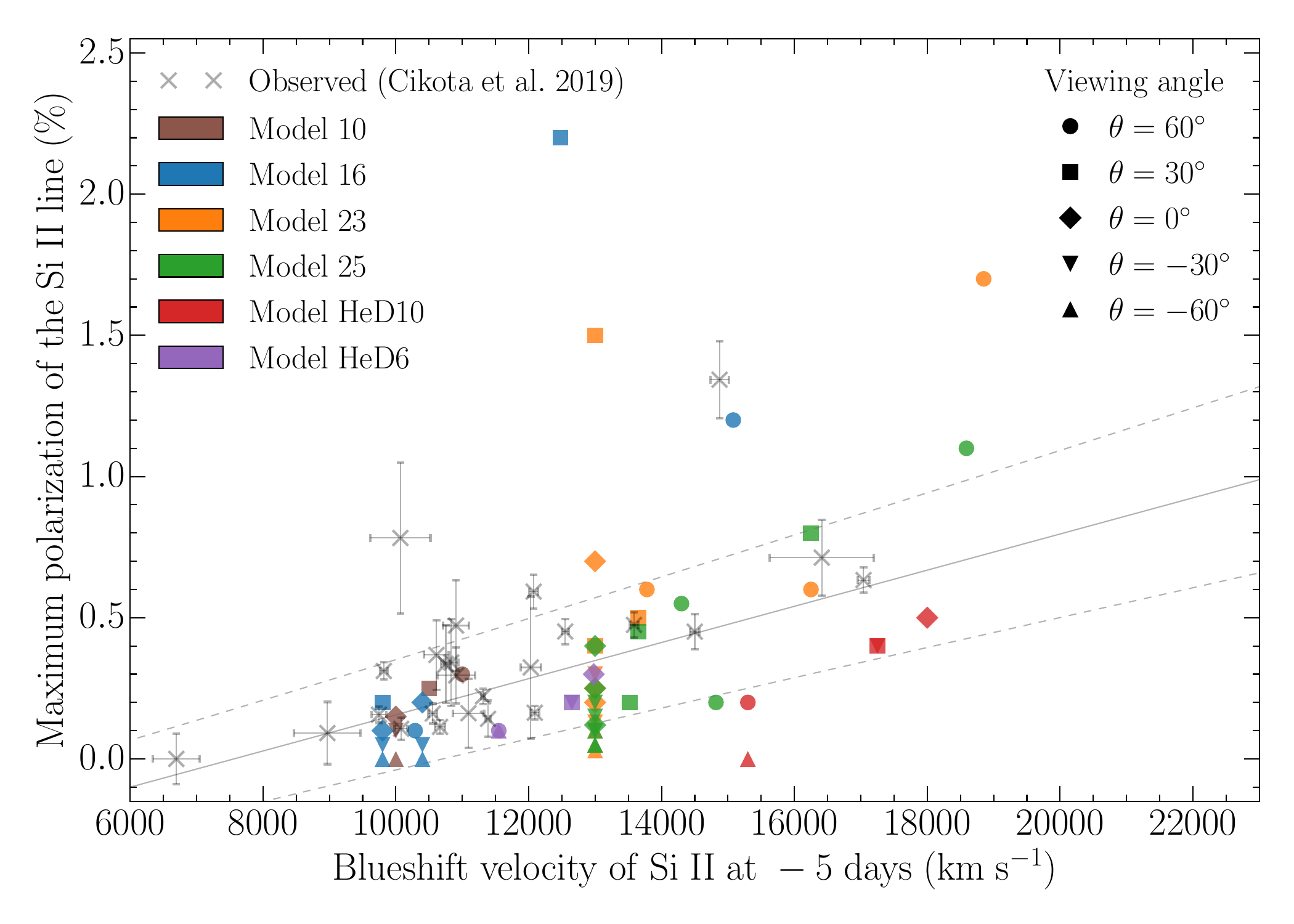}
\includegraphics[width=0.49\textwidth]{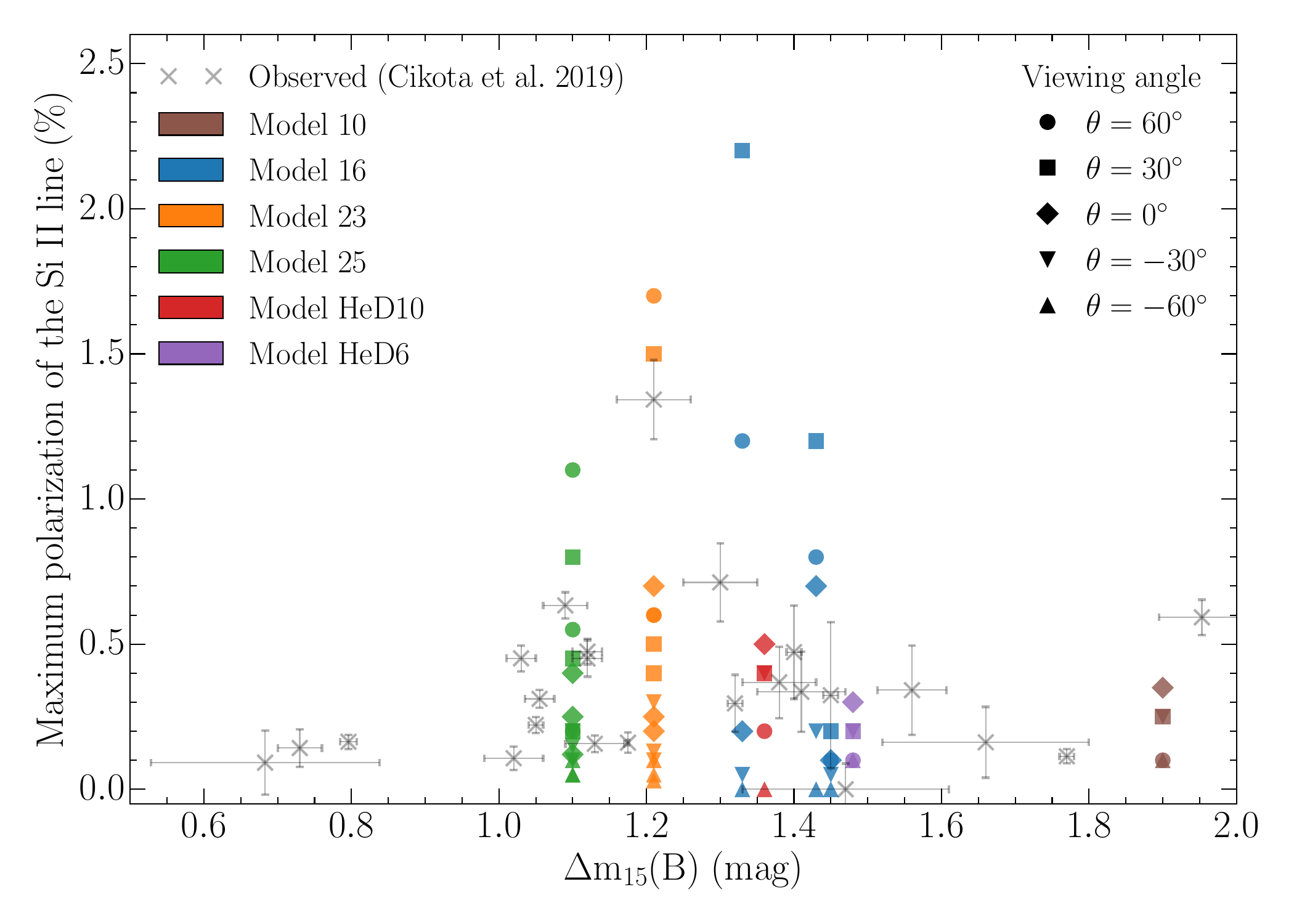}
\vspace{-3mm}
\caption{Comparison of the peak polarization in Si\,{\sc ii} as a function of the expansion velocity 5 days before peak brightness in the $B$ band (left panel), and the expansion velocity as a function of the luminosity decline rate (right panel), for a series of models from \citet{Hoeflich_2017}, compared to observations from \citet{Cikota2019MNRAS.490..578C}.
Model 10 assumes an elliptical oblate density structure with an axis ratio of A/B = 0.82 and reproduces subluminous SNe\,Ia. Models HeD10 and HeD6 represent sub-$\rm M_{Ch}$ SNe\,Ia, which include detonating He layers. 
Model 16 represents transitional SNe\,Ia, and models 23 and 25 represent normal-bright SNe\,Ia. 
Note that the polarization results for a viewing angle of +90 degrees are 0\%, and are not plotted. Such low polarization occurs only within a very narrow cone around the north pole ($\Delta \theta \approx 5^o$) and is expected in only about 1/50 SNe\,Ia. 
In the models, off-center delayed-detonation explosions viewed from angles near $\theta \rightarrow +90^\circ$ produce high expansion velocities for a given $\Delta$m$_{15}$ and low polarization (\PSi$\approx 0$).}
\label{fig:Models_PSiII_vel_asymmetry}
\end{figure*}

We measure the Si\,{\sc ii} expansion velocity from the absorption minima at 5 days before peak brightness in the modeled spectra. Note that the absorption minimum is only a first approximation \citep{1995ApJ...443...89H}, and its value depends on many variables, such as the main-sequence mass, the central density of the initial WD, and the metallicity.
This limits the accuracy of absolute values, but instead of focusing on absolute values, we focus on a qualitative analysis and investigate overall trends.
 For a wide variety of multi-dimensional simulations, the results are given in Table~\ref{table:simulationsresults}, and shown in Fig.\ \ref{fig:Models_PSiII_vel_asymmetry}.

Figure~\ref{fig:Models_PSiII_vel_asymmetry} (left) shows the maximum \PSi\ as a function of the Si\,{\sc ii} velocity at -5 days relative to the peak in B-band, compared to observations. Different colors mark different models, while the different symbols denote the viewing angle $\theta$. Models with low-velocity Si ejecta display low polarization values. In contrast, models with high velocity display higher polarization values. This trend is consistent with observations.
An exception occurs for models seen at a viewing angle of +90 degrees, which have low polarization values ($\sim$ 0\%) at high velocities. The reason is that if the model is viewed at +90 degrees (from the top in Fig.~\ref{fig:Schematics}), the ejected Si above the photosphere show an axisymmetric distribution along the line of sight. This produces zero net polarization because all Stokes vectors cancel out\footnote{The net cancellation of polarization requires that the scattering medium is unresolved on the sky, which is the case for all extragalactic supernovae observed with current UV/visible/IR telescopes.} (see also the discussion in Sect.~\ref{sect:directional_dependence}). In the simulations, the corresponding photon counter covers only 2\% of the total sky. Thus, only 1 out of 50 observed SNe is expected to fall in this cone. Moreover, in nature, such perfect axisymmetry is rare.
For both reasons, we do not observe low polarizations at high velocities for a given luminosity (i.e. $\Delta$m$_{15}$), but theoretically they are not impossible. 

The right panel of Fig.~\ref{fig:Models_PSiII_vel_asymmetry} shows the Si\,{\sc ii} velocity at 5 days before peak brightness in the B band as a function of $\Delta$m$_{15}$. As shown, subluminous SNe (with high $\Delta$m$_{15}$) tend to display low velocities and low polarization, while normal SNe (with normal $\Delta$m$_{15}$) tend to display higher velocities and higher degrees of polarization. This is also consistent with observations. 

For Class I models, to the first order, \PSi\ can be up to 1.5 or 2\% with a typical range between 0 and 0.7\% depending on the viewing angle (Fig.~\ref{fig:Models_PSiII_vel_asymmetry}). The spherical models from Table~\ref{table:simulationsresults} are not included in the figure because they do not display net polarization. 

Note that for the parameters considered here, the continuum polarization $\rm P_{cont}$ in the simulations is $\sim 0.1\%$ for Class I and $\sim 0.3\%$ for Class II asphericity, respectively. Throughout the paper, we subtract the continuum polarization and consider only the Si\,{\sc ii} polarization.

Normal-bright and transitional SNe are consistent with off-center DDT models 23 and 16, respectively. For off-center DDTs, the degree of \PSi\ depends sensitively on the location of the DDT relative to the transition between the layers of nuclear statistical equilibrium (Fe-group elements) and the incompletely burned oxygen layers. Consistent with detailed analyzes, clustering at $\rm M_{DDT}$ $\approx$ 0.3 $\rm M_{WD}$, well within the distributed burning regime \citep{2019Sci...366.7365P,2022APS..APRX13004U}, supports a turbulence-driven DDT. 
The polarization \PSi\ is highest at the lower-luminosity end of normal-bright and transitional SNe\,Ia (which also have high expansion velocities), both in observations and models (Table~\ref{table:simulationsresults}, Fig.~\ref{fig:Models_PSiII_vel_asymmetry}).
In off-center DDT scenarios, this can be understood because, at the epochs considered, velocities at the photosphere coincide with the same ejecta layers where the turbulence-driven DDT is expected to occur. 

Higher $\rm M_{DDT}$ are necessary to explain some cases with high \PSi, such as SN\,2004dt \citep{2006NewAR..50..470H} with $\rm M_{DDT}\approx$ 0.6 $\rm M_{WD}$; however, these are not common events within the framework of turbulence-driven DDTs.
Namely, out of the 35 SNe\,Ia with spectropolarimetry analyzed in \citet{Cikota2019MNRAS.490..578C}, SN\,2004dt is a clear outlier, exhibiting the highest Si\,{\sc ii} line polarization ever observed (1.34 $\pm$ 0.14 \%, and 1.37 $\pm$ 0.22 \% at -9.7 and 4.4 days relative to peak brightness, respectively; see the evolution of the Si\,{\sc ii} line polarization in Fig. 10 in \citealt{Cikota2019MNRAS.490..578C}, and Table A5, and Fig. B12 in the supplementary materials in \citealt{Cikota2019MNRAS.490..578C}). No other object in the sample reaches comparable levels, leading to the suggestion that such polarization degrees are intrinsically uncommon (see also \citealt{2007A&A...475..585A}).
However, the observed polarization depends on both the maximum polarization and the inclination angle, and SN\,2004dt may be the 'tip of an iceberg'. Fig.~\ref{fig:Models_PSiII_theta_asymmetry} suggests that the population of SN with large $M_{DDT}=0.6 M_\odot$ (red) represents a small fraction, approximately 15\%, and significantly contributes to the polarization distribution. Moreover, it is needed for \PSi\ $\geq$ 0.6 \%. This mix of populations is also consistent with the expansion velocities in Fig.~\ref{fig:Models_PSiII_vel_asymmetry}. Currently, our sample is too small to determine whether within $M_{DDT}$ a bimodal or continuous distribution is favorable, or alternatively, a completely different scenario from those discussed may manifest itself in SN\,2004dt without producing significant continuum polarization.



Although the overall trend of larger polarization with brightness and expansion velocity (Fig.~\ref{fig:PSiII_vel_asymmetry}) holds for most SNe\,Ia, the actual peak in \PSi\ occurs at the lower end of normal-bright and transitional SNe\,Ia in both off-center DDT models and observations.

Subluminous SNe with $\Delta$m$_{15}$ $>$ 1.6 are consistent with aspherical WDs or mergers with an axis ratio A/B $\sim$ 0.82, exemplified by model 10. This axis ratio was determined through detailed analyzes of individual supernovae. 
The polarization \PSi\ is smaller in subluminous SNe, because only $\approx 0.2 ~ \rm M_{WD}$ is undergoing burning to Nuclear Statistical Equilibrium (NSE) elements, and the amount of partially burned products such as S/Si is increased compared to bright SNe\,Ia (see Fig.\ 3 in \citealt{2002ApJ...568..791H}). Thus, the photosphere and Si line are mostly formed within the Si-rich region with a lesser contribution of the lopsided Si. This produces lower \PSi. As discussed above, the polarization spectra in subluminous SNe\,Ia such as SN\,1999by \citep{2001ApJ...556..302H} and SN\,2005ke \citep{2012A&A...545A...7P} require a significant aspheric component in the overall density distribution. Note that aspherical chemical distributions, e.g., due to Rayleigh-Taylor instabilities, inherently produce some amount of quasi-continuum polarization $\rm P_{cont}$ and asymmetric $^{56}$Ni and some Si\,{\sc ii} polarization by a large number of small-scale inhomogeneities produced, in particular, at the chemical interfaces between Si and NSE as observed in time-series of polarization spectra \citep{Yi_etal_2016,2023MNRAS.520..560H}.

For a given $\Delta$m$_{15}$, the highest velocities are observed at large viewing angles $\theta$ for Class I asphericities, and from the equator for Class II asphericities, respectively. For both asphericity classes and with any given brightness, high polarization is dominated by above-average photospheric expansion velocities.

For all scenarios, higher Si velocities imply more burning to NSE, but, generally, sub-$\rm M_{Ch}$ models show systematically higher velocities than near $\rm M_{Ch}$ explosions due to the lower binding energy of the WD with a similar specific nuclear energy release \citep{2002ApJ...568..791H,Shen_etal_2013,2017ApJ...838...36S}. 

As a result, high photospheric expansion velocities are dominated by bright supernovae (Fig.~\ref{fig:Models_PSiII_vel_asymmetry}). 
The expansion velocity is highest when viewed from the equator for aspheric envelopes and from the pole for off-center DDTs.
For off-center DDTs, both high expansion velocities and high Si\,{\sc ii} polarization values can be expected when viewed from a large $\theta \approx 60^o-80^o$.

\subsection{Relation between observed and intrinsic \PSi\ for large SNe samples}
\label{sect:4.2relatiobetween}

As a first step to link the statistical properties of models to observations, we show that the distribution of the observed \PSi\ in large samples of $n$ SNe is not affected by the direction from which an individual supernova has been observed\footnote{For SNe with multi-epoch observations, $\theta$ can be determined.}, and provide an estimate for the most likely intrinsic Si\,{\sc ii} polarization distribution, $P_{int}$, of the observed sample. 

For a given line of sight ($\theta$), the observed polarization is $P_{obs}= P_{int} \times f({\theta})$, where $P_{int}$ is the intrinsic polarization, and $f({\theta})$ is the probability of observing a SNe from a viewing angle $\theta$.
Although unlikely, $P_{int}$ could take arbitrarily large values for individual objects (up to 100\%). However, for a large sample of $n$ SNe, the uncertainty in
the average probability $\sum_{i=1}^n f(\theta_i)$ is $ 1/ \sqrt{n}$. 

\begin{figure}
\includegraphics[width=0.48\textwidth]{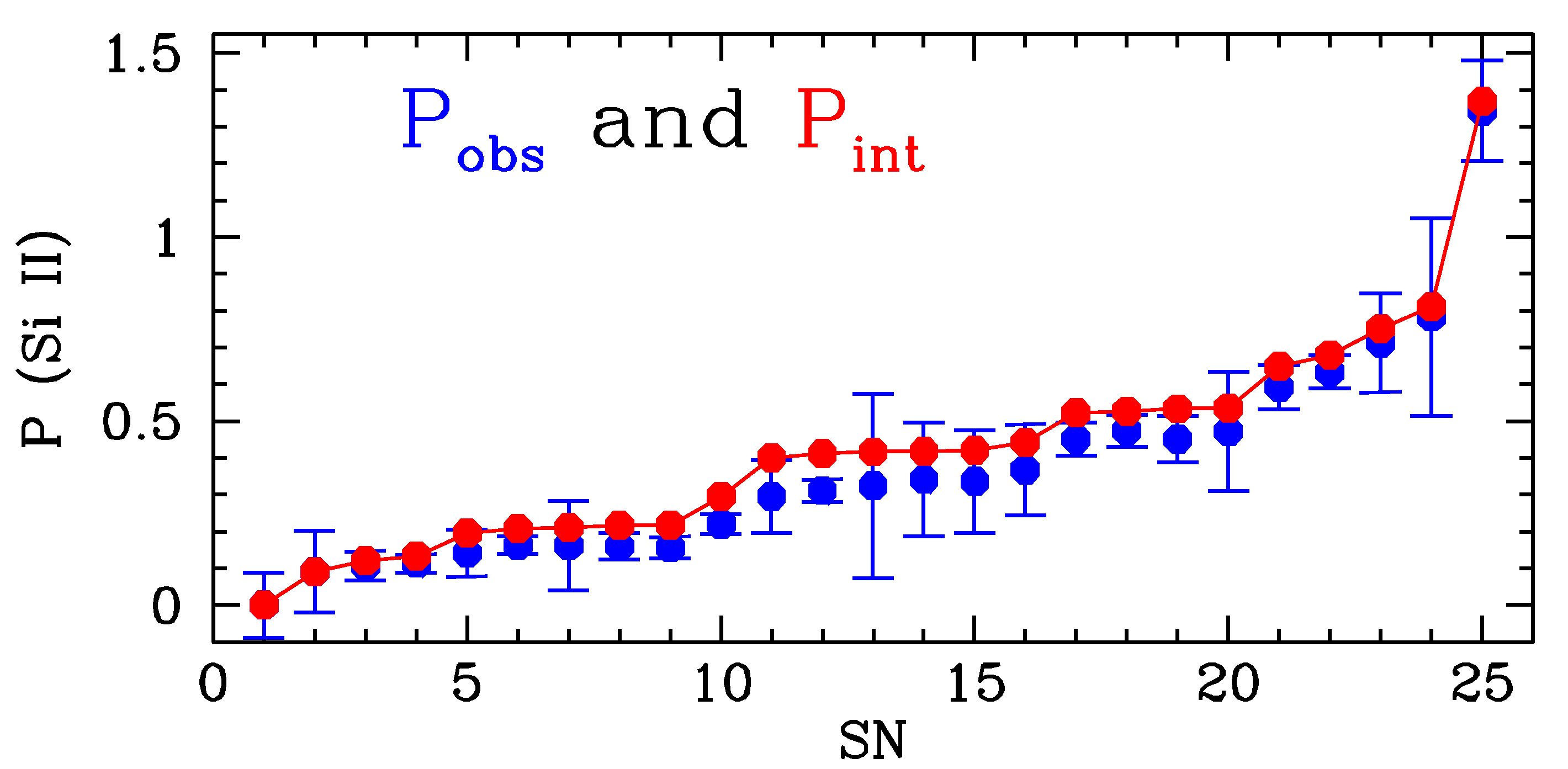}
\caption{Observed (blue) vs. intrinsic (red) polarization assuming maximum entropy for our SN sample assuming $\sim 0.1\%$ polarization (see text). The x-axis lists individual SNe\,Ia, sorted by their \PSi\ polarization degree. Note that the differences between the observed and intrinsic polarization degrees are small, comparable to the intrinsic errors and the assumed polarization due to small-scale clumps.}
\label{fig:PSiII_intr}
\end{figure}

Here, we estimate the most likely $n(P)$-distribution by maximizing the entropy of the sample (Fig.~\ref{fig:PSiII_intr}).   

Detailed analysis of simulations (Yi et al. 2025, in preparation) suggests that an additional \PSi\ $\approx 0.1 \%$ is produced by small-scale inhomogeneities (Class III). This assumption has been tested by assuming that the narrow cluster of normal-bright SNe at \PSi\ $\approx 0.3 \pm 0.1 \% $ has similar intrinsic polarization and that the variation is dominated by \PSi\ $(\theta)$ (see below).

\subsection{Viewing angle effects on the peak polarization}
\label{sect:viewingangleeffects}

The degree of line polarization in our models strongly depends on the orientation from which the SN is viewed. In this section, we investigate how different classes of asphericities imprint angle-dependent signatures on the Si\,{\sc ii} polarization. We first discuss the results of the full radiation-hydro models for Class I and II asymmetries, and then compare the predicted angle-dependent polarization distributions for different models with the available observations. Subsequently, employing a simplified model, we study the polarization arising from small-scale abundance clumps (Class III) with a large-scale distribution for the clumps. 

\label{sect:directional_dependence}
\subsubsection{Results of full radiation-hydro models}
\label{sect:resultsoffullratidaiton}

Figure~\ref{fig:Models_PSiII_theta_asymmetry} (left) shows the normalized Si\,{\sc ii} polarization distribution produced by chemical inhomogeneities in an off-center DDT (e.g.\ model 23 with $\rm M_{DDT}=0.3$, Class I asphericity) and ellipsoidal density structure (e.g.\ model HeD10, Class II asphericity), as a function of the viewing angle $\theta$. 

For a geometry that is rotationally symmetric in density (see e.g. Class II in Fig.~\ref{fig:Schematics}), \PSi($\theta$)=\PSi($-\theta$). In particular, \PSi\ is zero when the SN is viewed 'pole-on', because the underlying \PSi\ formed by Thomson scattering is zero. \PSi\ peaks when it is viewed from the equator. \PSi($\theta$) peaks because Si abundances follow the density distribution in both rapidly rotating WDs and mergers\footnote{Note that an aspheric density distribution also produces continuum polarization. For Class I asphericities, the continuum polarization remains small before maximum light because it is the result of the asymmetric energy input by radioactive decay of $\rm ^{56}Ni/^{56}Co$.} \citep{2001ApJ...556..302H,2012A&A...545A...7P}. 
Note that \PSi\ is not proportional to $\sin^2(\theta)$ as would be expected in the optical, Thomson-scattering-dominated case \citep{Van_de_Hulst_1957}
because, for low $\theta$, the photons thermalize at larger Thomson-scattering optical depths, and thus the effect of multiple-scattering becomes strongest (see, e.g., Fig.\ 3 in \citealt{1991A&A...246..481H}).

The off-center and lopsided configurations show a qualitatively different behavior (Fig.~\ref{fig:Schematics}, right panel). Prior to maximum light, the extended photosphere can obscure the Si-rich bulge. 
Hence, the Si line polarization depends strongly on the viewing angle and is no longer symmetric with respect to the equator.
The peak polarization arises for sight lines between +30 and +60 degrees, where the optical depth of Si becomes mostly uneven. This may also explain why detailed analyzes of the polarization spectra of well-observed SNe\,Ia with significant \PSi\ are biased toward positive $\theta$ (for example, SN\,2018gv and SN\,2019np by \citealt{Bufano_etal_2018} and \citealt{Hoeflich2023_2019np}, respectively). In contrast, the IR and MIR lines show no such bias (see Sect.~\ref{sect:simulations}).

Note that the distributions (Fig.~\ref{fig:Models_PSiII_theta_asymmetry}, left) are not necessarily related to the explosion mechanism. The Class I asphericity case represents any scenario that produces large-scale lopsided abundance distributions, whereas Class II asphericities may result from rapid rotation or dynamical mergers. Identification of the underlying physics of a specific object requires detailed analysis of time-dependent polarization spectra. 

\subsubsection{Comparison of the models to observations and discussion}
\label{sect:4.3.2 comparison}

\begin{figure*}
\centering
\includegraphics[width=0.98\textwidth]{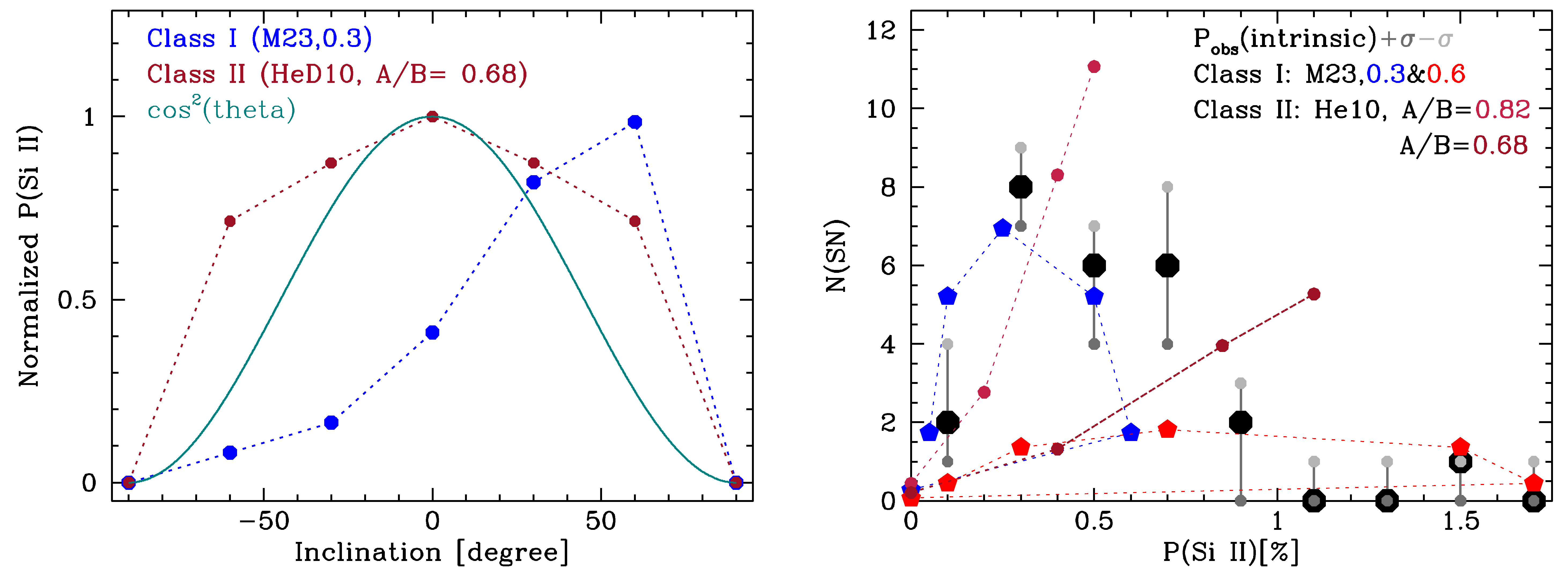}\caption{Viewing angle dependence of the \PSi\ polarization. \textit{Left:} Normalized Si\,{\sc ii} polarization (peak \PSi\ = 1) produced by abundance inhomogeneities in an off-center DDT model (Class I) vs.\ ellipsoidal density structure (Class II) as a function of viewing angle $\theta$. In addition, we give the likelihood $F\propto \cos^2(\theta)$ that a SN\,Ia is observed from a direction $\theta$.
As examples we show the normal-bright Model 23 with $\rm M_{DDT}\,=\,0.3\,M_{WD}$ \citep{2023MNRAS.520..560H} and the HeD10 model with an axis ratio of 0.68 (see Table~\ref{table:simulationsresults}). Note that all models of the same class have very similar \PSi$(\theta)$.
\textit{Right:} Distributions of the observed intrinsic maximum polarization (black dots, with observed uncertainties in N(SN) given in light and dark gray), the normal-bright models 23 with $\rm M_{DDT}\,=\,0.3\ \&\ 0.6\,M_{WD}$ (Class I, blue and red hexagons, respectively), and the normal-bright HeD10 with axis ratios of 0.82 and 0.68 (Class II, light and dark purple dots). The distribution of the off-center delayed-detonation models has been normalized to the total number of observations, with a fraction of $\approx 15 \%$ based on the high-polarization observations.}
\label{fig:Models_PSiII_theta_asymmetry}
\end{figure*}

Assuming a random and uniform distribution of the solid angle, i.e. of $\cos(\theta$), we determined the distribution of the maximum Si\,{\sc ii} polarization, shown in the form of histograms in Fig.~\ref{fig:Models_PSiII_theta_asymmetry} (right panel) for the different models and polarization dependencies. These distributions are compared to the observed distribution of the maximum Si\,{\sc ii} polarization measured over time (black dots). The values of the polarization depend on the model, for example, in Model 23, for $\rm M_{DDT}$ = 0.3 $\rm M_{WD}$, the polarization of Si\,{\sc ii} is \PSi\ = 0.6\%, and for $\rm M_{DDT}$ = 0.6 $\rm M_{WD}$, \PSi\ = 1.7\% (see Table~\ref{table:simulationsresults}). Despite these differences in amplitude, the overall shape of the distributions is qualitatively similar.


We show in Fig.~\ref{fig:Models_PSiII_theta_asymmetry} (right panel) that the off-center or lopsided configurations (Class I) have a flat distribution, with an increased frequency of low polarization values \PSi$\lesssim 0.5\%$, 
The observed sample also shows a high frequency of SNe with low polarization values in Si\,{\sc ii} (\PSi$\lesssim 0.5\%$), with the frequency decreasing approximately linearly toward higher values. 
However, we estimated that about $\sim 15\%$ of the SNe require models with $\rm M_{DDT} \approx 0.6~ M_{WD}$ to account for the high polarization values observed at the low-luminosity end of normal-bright SNe\,Ia (Fig.~\ref{fig:Models_PSiII_theta_asymmetry}). Note that the high polarization observed in transitional SNe\,Ia (e.g.\ SN\,2003W, \citealt{Cikota2019MNRAS.490..578C} is consistent with $\rm M_{DDT}\approx 0.3~M_{WD}$ (Model 16, see Table~\ref{table:simulationsresults} and Fig.~\ref{fig:Models_PSiII_vel_asymmetry}). This spread in $\rm M_{DDT}$ may suggest two different mechanisms for a DDT.

The observational dataset is more consistent with off-center or lopsided configurations than with the models representing Class II asphericities (see black dots compared to blue and red in the right panel of Fig.~\ref{fig:Models_PSiII_theta_asymmetry}).
Note that for the HeD models (Class II), at a given $A/B$, the total polarization is higher because both the density and abundances are aspherical resulting in the high polarization in Si compared to nearly spherical density distribution. The line opacity reduces the non-canceled polarization by Thomson scattering. Because both abundances and density have the same axis, this leads to higher \PSi\ \citep{2012ApJ...747L..10P}. 
The observations of Si\,{\sc ii} polarization cannot be reproduced by Class II without fine-tuning A/B. Although possible, such asymmetries are accompanied by high continuum polarization, which, as discussed from the detailed analysis of SNe\,Ia (see Sect.~\ref{sect:introduction}), can be excluded for normal-bright and transitional SNe\,Ia, but is favored in underluminous SNe\,Ia.

\subsubsection{A simple model: Plumes (Class III) above a well-defined Thomson-scattering photosphere}
\label{sect:model_classIII}

In this section, we consider the case in which the photosphere is not offset nor ellipsoidal, and the polarization is caused by a large-scale aspherical distribution of Si\,{\sc ii} clumps (Class III asphericity) on top of a sharp photosphere. We employ a simplified model with the number of clumps N as a free parameter. For simplification, we neglect multiple scattering effects within the clumps, assume that each plume has the same optical depth, and neglect that lines and continuum polarization are formed in the same layers \citep{Hoeflich2023_2019np}.

For a given large-scale asymmetry, the degree of polarization depends in a complex way on the size and number of clumps and can be explored with a Monte Carlo simulation, as shown in \citet{2007Sci...315..212W}.
 Polarization can be modeled by a random walk of vectors of N steps, each of length $\sim$10\% along the edge of the photospheric disk (Lifan Wang, unpublished; Yang et al., in prep.). 
Polarization reduces to zero if N is infinite, so a moderate number of large clumps are most likely to produce a high degree of polarization.

For an assumed angle-dependent distribution of large-scale clumps, an arbitrary angle-dependent distribution of polarization can be achieved.
The polarization dependency on the viewing angle may show a double-peaked profile given by $\rm cos^2(\theta)$ for a globally prolate Si\,{\sc ii} asymmetry and $\rm sin^2(\theta)$ for an oblate Si\,{\sc ii} asymmetry. 

In the case of large-scale distributions of clumps as a function of polar angle, net depolarization is proportional to $\rm \sqrt{N}$, so the degree of polarization could be proportional to $\rm 1-\sqrt{N(\theta)}/N(\theta) = 1-1/ \sqrt{N}$ in a random walk scenario. If we assume that N($\theta$) is proportional to $\rm 1/cos(\theta)$, with the clumps smaller at the equator than at the poles, the polarization dependency may show a profile given by $1 -\sqrt{\cos(\theta)}$, which would produce an observed \PSi\ distribution (e.g.\ in the right panel of Fig.~\ref{fig:Models_PSiII_theta_asymmetry}) with a high frequency of SNe with small Si\,{\sc ii} polarization values, linearly decreasing towards higher values. 

We note that Rayleigh-Taylor instabilities \citep[e.g.][]{Gamezo_etal_2004,2012A&A...545A...7P} or cellular instabilities \citep{2025ApJ...982..204K}, do not produce any angle-dependence of the polarization but, if individually unresolved in simulations (see Sect.~\ref{sec:numerical_tools}), instead produce a scatter in the $Q$-$U$ plane.

\section{Summary and conclusions}
\label{sect:summary}

We examined a sample of off-center delayed-detonation models with different deflagration-to-detonation transition masses. The models represent lopsided abundance distributions (Class I asphericities) that reproduce normal-bright SN\,Ia observations, and aspherical oblate models with elliptical density and abundance structures with or without a He layer (Class II asphericities) that reproduce subluminous SN\,Ia observations \citep{hk96,2001ApJ...556..302H,2012A&A...545A...7P,Hoeflich_2017}.
In the modeled spectra, we measured the maximum polarization of Si \,{\sc ii} $\rm \lambda 6355\,\AA$ (with respect to time) and the expansion velocity of Si \,{\sc ii} 5 days before the peak brightness of the B-band, and compared the results with observations from \citet{Cikota2019MNRAS.490..578C}. 

The main findings can be summarized as follows:\\
(i) The off-center delayed-detonation simulations show a correlation between Si\,{\sc ii} polarization and expansion velocity, with higher velocities leading to higher polarization values, consistent with observations (Fig.~\ref{fig:Models_PSiII_vel_asymmetry} left panel). This can be understood as an orientation effect: higher polarization is obtained when the models are viewed from positive $\theta$ (see Fig.~\ref{fig:Schematics}). Consequently, within every group of $\Delta$m$_{15}$, \PSi\ correlates with $v$ (see Sect.~\ref{sect:max_lin_pol_v_dm15}). Note that, within this scenario, the nature of the high-velocity SNe may be understood as an orientation effect (Fig.~\ref{fig:Models_PSiII_vel_asymmetry}, left panel). \\
(ii) We found that the polarization of Si\,{\sc ii} in normal-bright SNe\,Ia is dominated by the aspherical abundances produced during explosive oxygen burning, while in underluminous SNe\,Ia, large-scale density distributions dominate. For subluminous SNe\,Ia, Class II asymmetries dominate (see Sect.~\ref{sect:max_lin_pol_v_dm15}).\\ 
(iii) From simulations, off-center delayed-detonations cover the entire range of \PSi\ observed with the highest polarization (up to 2\%) occurring at the lower end of normal-bright to transitional SNe (Fig.~\ref{fig:Models_PSiII_vel_asymmetry}).
For DDT in the turbulence-driven regime, the typical \PSi\ is about 0.5\%, suggesting that this is the dominant mechanism. Within DDT models, higher \PSi\ indicates that larger $\rm M_{DDT}$ are realized in rare cases, potentially hinting at multiple DDT mechanisms.\\
(iv) Within the DDT scenario and for normal-bright and transitional SNe, the location $\rm M_{DDT}$ for the delayed-detonation transition is consistent with the turbulence-driven DDT mechanism (see Sect.~\ref{sect:results}, Fig.~\ref{fig:Models_PSiII_theta_asymmetry}, right). Although rare, more off-center DDTs are needed to reproduce objects with large \PSi, suggesting that alternative mechanisms for DDT are possible \citep{1995ApJ...452..779N, 1997ApJ...478..678K, 2025ApJ...982..204K,Livne_1999,2003ASPC..288..185H}.\\
(v) \PSi\ peaks at the lower luminosity of normal-bright and transitional SNe\,Ia in both observations and models (Fig.~\ref{fig:Models_PSiII_vel_asymmetry}, right panel). Within the off-center DDT models, this can be understood because, at the times considered, the velocities at the photosphere coincide with those same layers in which the turbulence-driven DDT is expected (see the right panel of Fig. \ref{fig:Schematics}). \\
(vi) Subluminous SNe have polarization properties consistent with large-scale rotationally axisymmetric asphericities in the density distribution caused by the double-degenerate channel or a rapidly rotating WD (Class II), as suggested in a few cases by detailed analyzes of specific objects \citep{2001ApJ...556..302H,2012A&A...545A...7P}.
As discussed in Sect.~\ref{sect:simulations}, Class II asphericities may be caused by other explosion scenarios.\\
(vii) However, we emphasize the ubiquity of persistent loops in the $Q$-$U$ plane \citep[e.g.][]{Cikota2019MNRAS.490..578C} strongly suggests the presence of, at least, two axes produced, e.g.\ by rotational symmetry in the density and by lopsided abundances. 
Because \PSi\ is produced predominantly by Thomson scattering in the extended line-forming region, from a theoretical point of view, one may think of a combination of the two cases as a linear overlap of Classes I and II with the size of the loops as a measure of the relative importance of the components. 
Note that loops may also be produced by small-scale asphericities as expected from shear and Rayleigh-Taylor instabilities (Class III), which, for proper distinction, require time-sequence observations (see Sects.~\ref{sect:4.2relatiobetween} and \ref{sect:4.3.2 comparison}).\\
(viii) We compared the polarization distributions of Si\,{\sc ii} with simplified assumptions associated with Class III asphericities. Our Class III ``models'' are not as developed as those presented for Classes I and II. This prevents detailed studies, such as the reproduction of the velocity–polarization relationship. However, we note that the observational data may also be reproduced by a configuration with a spherical photosphere and a clumpy, globally prolate Si\,{\sc ii} distribution, or alternatively a turbulent structure, rather than an off-center geometry (see Sect.~\ref{sect:model_classIII}).\\ 

Understanding the explosion physics and their systematics is critical, because our results may have relevance for high-precision SN Ia cosmology. Our models applied to local SNe~Ia imply a viewing-angle dependence of the luminosity of $\approx 2 - 3 \% $ for the average luminosity observed \citep{1991A&A...246..481H,2001ApJ...556..302H,2006NewAR..50..470H,2012A&A...545A...7P,2020ApJ...902...46Y,Hoeflich2023_2019np}. Note that a corresponding uncertainty in the distance may arise when using statistical means to obtain the high accuracy needed to probe deviations from standard cosmology, as implied by the DESI collaboration \citep{2026JHEAp..4900428O}.

We demonstrated the effectiveness of multi-object, single-snapshot spectropolarimetry during the pre-maximum phase by combining it, for the first time, with a statistical analysis that links theoretical model distributions to observed polarization data. 
This approach provides an efficient means to probe the diversity of SNe~Ia, offering strong complementarity to time-series observations that reveal the underlying 3D physics.

It is essential to acknowledge the limitations of this study. 
We used off-center DDT explosions because their energy input is centrally concentrated. Other scenarios, such as WD collisions, would also show a strong offset in the energy input but, at early times, similar Si\,{\sc ii} polarization.
However, in the case of WD collisions, the polarization would show a flip in sign over time, which has not been observed in any spectropolarimetric observation of SNe\,Ia \citep{1995ApJ...443...89H,Bulla_etal_2018}. 

We have also not extensively considered small-scale asphericities such as clumps (Class III), which will be the subject of future analyzes and will require low-latency time-sequence observations. 
Moreover, we have not conducted a quantitative analysis of detailed 3D abundance structures for sub-Chandrasekhar mass explosions using non-LTE simulations, due to the broad diversity of possible scenarios. As an initial step, identifying the viable parameter space through a multi-wavelength approach is essential, followed by polarization simulations. Work in this direction is currently underway.

\bigskip

{\sl Acknowledgments} 
The authors thank all colleagues of the SPECPOL team and collaborators who over the years contributed to the database and modeling efforts.
The work of A.C. is supported by NOIRLab, which is managed by the Association of Universities for Research in Astronomy (AURA) under a cooperative agreement with the National Science Foundation. L.W. is partially supported by the NSF through the grant AST 1813825. 
P.H. and the simulations in this paper are partially supported by the NSF through grants AST 171513 and 230639, and NASA grants JWST-GO-02114, GO-02122, GO-03727, GO-004217, GO-04436, GO-4522, GO-5057, GO-05290, GO-06023, GO-6677.
This work is based on observations collected at the European Organization for Astronomical Research in the Southern Hemisphere under ESO programs 67.D-0517(A), 66.D-0328(A), 68.D-0571(A), 69.D-0438(A), 70.D-0111(A), 71.D-0141(A), 073.D-0771(A), 073.D-0565(A), 075.D-0628(A), 075.D-0213(A), 076.D-0178(A), 076.D-0177(A), 079.D-0090(A), 080.D-0108(A), 081.D-0557(A), 081.D-0558(A), 085.D-0731(A), 086.D-0262(A), 086.D-0262(B), 088.D-0502(A), 095.D-0848(A), 095.D-0848(B), 290.D-5009(A), 290.D-5009(B), 290.D-5009(C), and 290.D-5009(D); the execution in service mode of these observations by the VLT operations staff is gratefully acknowledged.

\bigskip
\noindent
{\sl Facilities:}
The observations were obtained with the Very Large Telescope at the European Southern Observatory's La Silla Paranal Observatory in Chile. The simulations have been performed on the computer cluster of the astro-group at Florida State University. 

\bigskip
\noindent
{\sl Software:}
IRAF is distributed by NOIRLab, which is operated by the Association of Universities for Research in Astronomy, Inc., under cooperative agreement with the National Science Foundation. The HYDRA code and various related modules were used in this work.

\bigskip
\noindent
{\sl Data Availability Statement:} 
All data supporting this work is publicly available. The raw FORS observations can be accessed through the ESO Science Archive (https://archive.eso.org/) using the corresponding Program IDs given in the acknowledgments.

\bibliography{references,bibtex_pah}{} 



\end{document}